\RequirePackage{fix-cm}

\documentclass[smallcondensed,a4paper]{svjour3}

\smartqed
\usepackage{graphicx}

\begin{document}

\title{Superconductivity in Al-Nb-Ti-V-Zr multicomponent alloy}

\titlerunning{Superconductivity in Al-Nb-Ti-V-Zr multicomponent alloy}

\author{Yuta Harayama         \and
        Jiro Kitagawa
}

\institute{Yuta Harayama  \at
              Department of Electrical Engineering, Faculty of Engineering, Fukuoka Institute of Technology, 3-30-1 Wajiro-higashi, Higashi-ku, Fukuoka 811-0295, Japan \\
           \and
           Jiro Kitagawa \at
             Department of Electrical Engineering, Faculty of Engineering, Fukuoka Institute of Technology, 3-30-1 Wajiro-higashi, Higashi-ku, Fukuoka 811-0295, Japan \\
           \email{j-kitagawa@fit.ac.jp}
}

\date{Received: date / Accepted: date}

\maketitle

\begin{abstract}
The superconducting high-entropy alloys (HEAs) recently attract considerable attention due to their exciting properties, such as the robustness of superconductivity against atomic disorder and extremely high-pressure. The well-studied crystal structure of superconducting HEAs is body-centered-cubic (bcc) containing Nb, Ti, and Zr atoms. The same elements are contained in Al$_{5}$Nb$_{24}$Ti$_{40}$V$_{5}$Zr$_{26}$, which is a recently discovered bcc HEA and shows a gum-metal-like behavior after cold rolling. The gum metal is also an interesting system, exhibiting superelasticity and low Young's modulus. If gum metals show superconductivity and can be used as a superconducting wire, the gum-metal HEA superconductors might be the next-generation superconducting wire materials. Aiming at a fundamental assessment of as-cast Al-Nb-Ti-V-Zr multicomponent alloys including Al$_{5}$Nb$_{24}$Ti$_{40}$V$_{5}$Zr$_{26}$, we have investigated the structural and superconducting properties of the alloys. All alloys investigated show the superconductivity, and the valence electron concentration dependence of the superconducting critical temperature is very close to those of typical superconducting bcc HEAs.
\end{abstract}

\section{Introduction}
High-entropy alloy (HEA) is a new class of materials and was initially proposed for simple crystal structures such as face-centered-cubic, body-centered-cubic (bcc), and hexagonal-closed packing in which more than five elements, each having the atomic fraction between 5\% and 35\%, randomly occupy one crystallographic site\cite{Gao:book,Murty:book}.
Nowadays, the HEA concept is adopted in many alloys with multiple Wyckoff positions, oxides, chalcogenides, etc.\cite{Kitagawa:Metals,Musico:APLMat,Jiang:Science}.
The definition of HEA other than by the atomic concentration relies on the order of mixing entropy $\Delta S_\mathrm{mix}$: low-entropy alloys as having $\Delta S_\mathrm{mix}$ less than 0.69$R$ ($R$ is the gas constant), medium-entropy alloys with $\Delta S_\mathrm{mix}$ between 0.69$R$ and 1.60$R$, and HEAs with $\Delta S_\mathrm{mix}$ of 1.60$R$ or larger\cite{Otto:ActaMater}.
The high atomic-disorder leads to the severe lattice distortion effect, and especially bcc HEAs show the enhancement of mechanical properties, which attract researchers all over the world.
The enhancement of property beyond the simple mixture of those of constituent elements is also observed as, for example, outstanding thermal stability or enhanced magnetic frustration\cite{Sun:MRL,Marik:ScrMat}.

The HEA superconductivity is one of the hot topics of the HEA research field due to its exotic properties\cite{Kitagawa:Metals,Sun:PRM}.
For example, the superconductivity is robust against atomic disorder and highly insensitive to the application of extremely high-pressure\cite{Rohr:PNAC,Guo:PNAC}.
Although the bcc HEA superconductors have been well studied\cite{Rohr:PNAC,Guo:PNAC,Kozelj:PRL,Rohr:PRM,Marik:JALCOM,Ishizu:RINP,Nelson:SciRep,Zhang:PRR,Kim:ActaMater}, several HEAs with multisite crystal structures such as the CsCl-type, A15, NaCl-type, $\sigma$-phase, and layered structures also show the interesting superconductivities\cite{Stolze:ChemMater,Wu:SCM,Mizuguchi:JPSJ,Yamashita:DalTra,Liu:JALCOM,Sogabe:APE,Sogabe:SSC}.
While the relation between the superconductivity and the high-entropy state is unclear in many systems, in high-entropy (La$_{0.2}$Ce$_{0.2}$Pr$_{0.2}$Nd$_{0.2}$Sm$_{0.2}$)O$_{0.5}$F$_{0.5}$BiS$_{2}$ with the BiS$_{2}$-based superconducting layer and the REO (RE: rare earth) blocking layer, the bulk nature of superconductivity is improved compared to PrO$_{0.5}$F$_{0.5}$BiS$_{2}$ with the low-entropy state, which suggests that the mixing entropy at the blocking layer severely affects the superconducting state\cite{Sogabe:APE,Sogabe:SSC}.

The gum metals, which have received considerable attention due to their unusual mechanical behaviors such as superelasticity and low Young's modulus\cite{Saito:Science}, possess chemical compositions similar to those of bcc HEA superconductors.
The unusual properties are obtained after the cold rolling of as-cast materials.
The valence electron concentration per atom (VEC) of the gum metals is 4.24, which is appropriate for the appearance of the superconductivity of $d$-electron alloy superconductors.
Recently Al$_{5}$Nb$_{24}$Ti$_{40}$V$_{5}$Zr$_{26}$ is reported as a gum-metal-like HEA\cite{Zherebtsov:Int}.
If gum metals show superconductivity, they are highly advantageous for making superconducting wires, and this kind of material may be a good candidate for the next-generation superconducting wire.

In this paper, we have started from Al$_{5}$Nb$_{24}$Ti$_{40}$V$_{5}$Zr$_{26}$, and characterized the structural and superconducting properties of Al$_{5}$Nb$_{x}$Ti$_{35}$V$_{5}$Zr$_{55-x}$ and Al$_{5}$Nb$_{x}$Ti$_{59-x}$V$_{5}$Zr$_{31}$ alloys in the as-cast form.
The superconducting properties were investigated by measuring the ac magnetic susceptibility and the electrical resistivity.
This study will contribute to the basic understanding of superconductivity in gum metals in the future.

\section{Materials and methods}
Eight polycrystalline samples with different atomic compositions, as listed in Table 1, were synthesized by a home-made arc furnace.
The constituent elements of Al (99.99 \%), Nb (99.9 \%), Ti (99.9 \%), V (99.9 \%), and Zr (99.5\%), were arc-melted on a water-cooled Cu hearth in an Ar atmosphere.
The samples received no heat treatment.

The X-ray diffraction (XRD) patterns of prepared samples were checked by an X-ray diffractometer (XRD-7000L, Shimadzu, Kyoto, Japan) with Cu-K$\alpha$ radiation.
We used thin slabs cut from the as-cast samples due to their high ductility.
The microstructure of each sample was investigated by a field emission scanning electron microscope (FE-SEM, JSM-7100F; JEOL, Akishima, Japan).
The atomic compositions of the samples were determined by an energy dispersive X-ray (EDX) spectrometer equipped to the FE-SEM.

The temperature dependence of ac magnetic susceptibility $\chi_{ac}$ ($T$) in an alternating field of 5 Oe at 800 Hz, between 2.8 K and 300 K, was measured by a mutual inductance method using a GM refrigerator (UW404, Ulvac cryogenics, Kyoto, Japan).
The temperature dependence of electrical resistivity $\rho$ ($T$) between 2.8 K and 300 K was measured by the conventional dc four-probe method using the GM refrigerator.

\begin{table}
\caption{Atomic composition determined by EDX measurement, lattice parameter $a$, $\delta$, and $\Delta S_\mathrm{mix}/R$ of prepared samples.}
\label{t1}
\begin{tabular}{ccccc}
\hline
Sample & Atomic composition & a (\AA) & $\delta$ & $\Delta S_\mathrm{mix}/R$ \\
\hline
Al$_{5}$Nb$_{24}$Ti$_{40}$V$_{5}$Zr$_{26}$ & Al$_{5.1(3)}$Nb$_{25(1)}$Ti$_{38.6(6)}$V$_{4.6(2)}$Zr$_{27(1)}$ &3.349(3) & 5.28 & 1.359 \\
Al$_{5}$Nb$_{14}$Ti$_{35}$V$_{5}$Zr$_{41}$ & Al$_{5.1(3)}$Nb$_{14.6(5)}$Ti$_{33.6(3)}$V$_{4.9(4)}$Zr$_{41.8(4)}$ &3.407(4) & 5.72 & 1.308 \\
Al$_{5}$Nb$_{24}$Ti$_{35}$V$_{5}$Zr$_{31}$ & Al$_{4.8(2)}$Nb$_{25.7(7)}$Ti$_{33.6(3)}$V$_{4.5(5)}$Zr$_{31.4(3)}$ &3.368(1) & 5.53 & 1.373 \\
Al$_{5}$Nb$_{34}$Ti$_{35}$V$_{5}$Zr$_{21}$ & Al$_{5.0(1)}$Nb$_{33.5(6)}$Ti$_{33.5(3)}$V$_{5.2(3)}$Zr$_{22.9(5)}$ &3.342(2) & 5.06 & 1.362 \\
Al$_{5}$Nb$_{44}$Ti$_{35}$V$_{5}$Zr$_{11}$ & Al$_{4.6(2)}$Nb$_{45.1(8)}$Ti$_{34.1(3)}$V$_{4.7(1)}$Zr$_{11.5(7)}$ &3.314(2) & 4.18 & 1.271 \\
Al$_{5}$Nb$_{14}$Ti$_{45}$V$_{5}$Zr$_{31}$ & Al$_{4.7(2)}$Nb$_{12.9(4)}$Ti$_{47.7(5)}$V$_{5.5(3)}$Zr$_{29.2(5)}$ &3.372(1) & 5.40 & 1.297 \\
Al$_{5}$Nb$_{34}$Ti$_{25}$V$_{5}$Zr$_{31}$ & Al$_{4.9(2)}$Nb$_{32.7(5)}$Ti$_{27.7(3)}$V$_{6.0(4)}$Zr$_{28.7(7)}$ &3.371(2) & 5.65 & 1.376 \\
Al$_{5}$Nb$_{44}$Ti$_{15}$V$_{5}$Zr$_{31}$ & Al$_{5.0(2)}$Nb$_{40.9(7)}$Ti$_{17.6(2)}$V$_{5.8(3)}$Zr$_{30.6(7)}$ &3.371(1) & 5.77 & 1.308 \\
\hline
\end{tabular}
\end{table}

\section{Results and discussion}
The XRD pattern of Al$_{5}$Nb$_{24}$Ti$_{40}$V$_{5}$Zr$_{26}$ is characterized by a single-phase bcc with the lattice parameter $a=$3.349 $\AA$, which is consistent with the reported value 3.350 $\AA$\cite{Zherebtsov:Int} (see Table 1 and the inset of Figure 1(a)).
Al$_{5}$Nb$_{24}$Ti$_{40}$V$_{5}$Zr$_{26}$ exhibits the diamagnetic signal and the zero resistivity as shown in Figs.\ 1(a) and 1(b), respectively.
In the measurement of the real part of ac magnetic susceptibility $\chi_{ac}^{'}$ ($T$), the superconducting critical temperature $T_\mathrm{c}$ is determined as being the intercept of the linearly extrapolated diamagnetic slope with the normal state signal\cite{Rohr:PNAC,Hamamoto:MRX} (see the broken lines in Fig.\ 1(a)).
$T_\mathrm{c}$ in $\rho$ ($T$) measurement is determined from the onset of the transition.
The inset of Fig.\ 1(b) shows the $\rho$ data in the whole measured temperature range, which indicates the weak temperature dependence in the normal state due to the atomic disorder.

\begin{figure}
\begin{center}
\includegraphics[width=\linewidth]{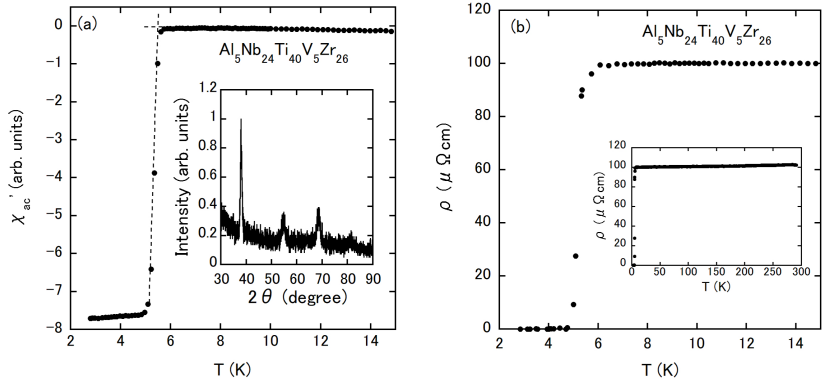}
\end{center}
\caption{(a) Temperature dependence of real part of $\chi_{ac}$ for Al$_{5}$Nb$_{24}$Ti$_{40}$V$_{5}$Zr$_{26}$. The inset is the XRD pattern of Al$_{5}$Nb$_{24}$Ti$_{40}$V$_{5}$Zr$_{26}$. (b) Temperature dependence of $\rho$ for Al$_{5}$Nb$_{24}$Ti$_{40}$V$_{5}$Zr$_{26}$ at low temperatures. The inset is $\rho$ in the whole measured temperature range.}
\label{f1}
\end{figure}

\begin{figure}
\begin{center}
\includegraphics[width=\linewidth]{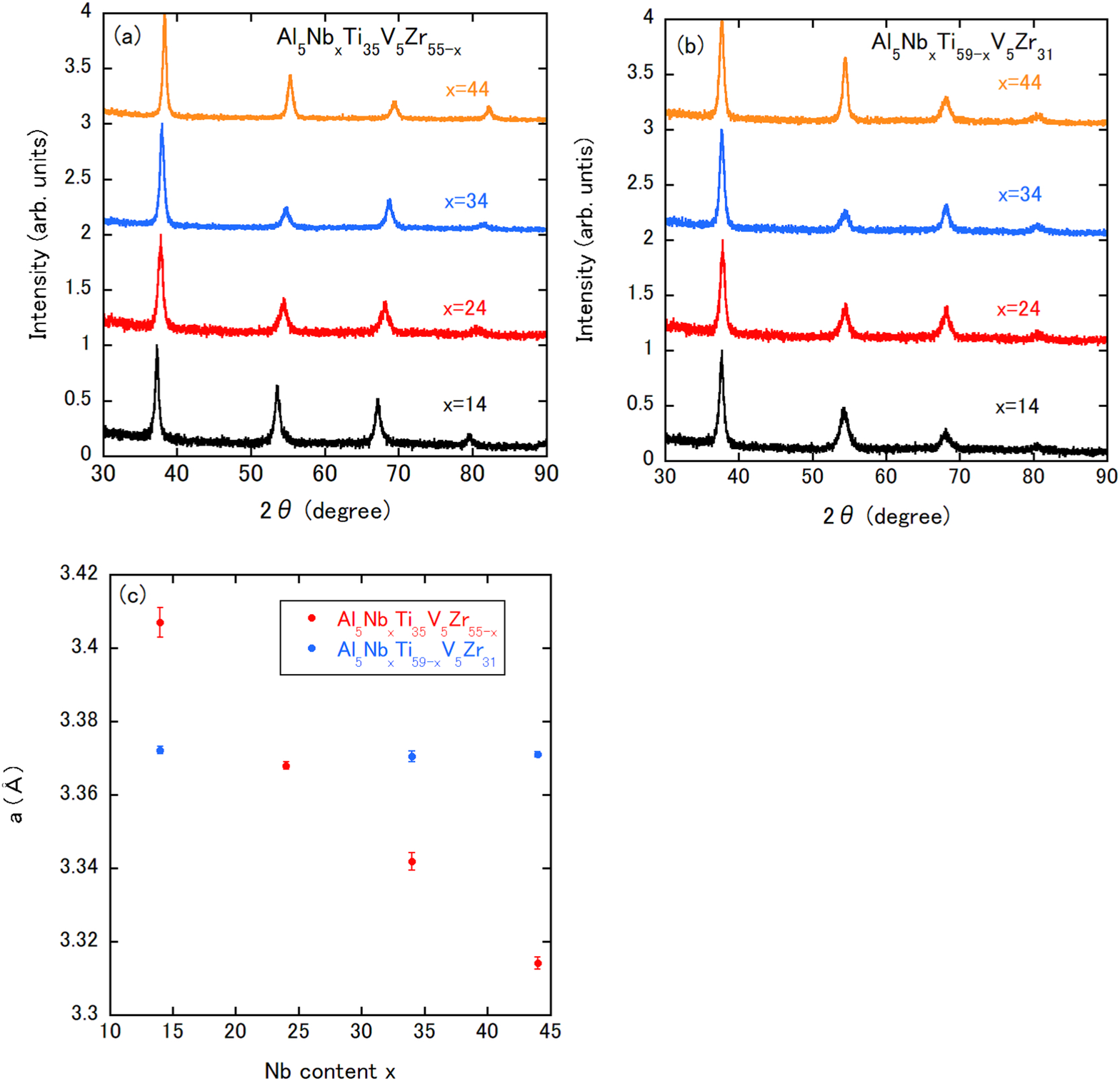}
\end{center}
\caption{X-ray diffraction patterns of (a) Al$_{5}$Nb$_{x}$Ti$_{35}$V$_{5}$Zr$_{55-x}$ and (b) Al$_{5}$Nb$_{x}$Ti$_{59-x}$V$_{5}$Zr$_{31}$, respectively. The origin of each pattern is shifted by an integer value for clarity. (c) Nb content dependences of lattice parameter of Al$_{5}$Nb$_{x}$Ti$_{35}$V$_{5}$Zr$_{55-x}$ and Al$_{5}$Nb$_{x}$Ti$_{59-x}$V$_{5}$Zr$_{31}$.}
\label{f2}
\end{figure}

Figures 2(a) and 2(b) show the XRD patterns of Al$_{5}$Nb$_{x}$Ti$_{35}$V$_{5}$Zr$_{55-x}$ and Al$_{5}$Nb$_{x}$Ti$_{59-x}$V$_{5}$Zr$_{31}$, respectively.
In each sample, the diffraction peaks can be well indexed by a bcc structure.
The lattice parameters are obtained by a least-mean square method (see Table 1) and plotted as a function of Nb content $x$ in Fig.\ 2(c).
For Al$_{5}$Nb$_{x}$Ti$_{35}$V$_{5}$Zr$_{55-x}$, the lattice parameter shrinks with increasing $x$, which is contrasted with the almost $x$-independent lattice parameters in Al$_{5}$Nb$_{x}$Ti$_{59-x}$V$_{5}$Zr$_{31}$.
The SEM images of representative samples are displayed in Figures 3(a) and 3(b), indicating no apparent impurity phase.
Each elemental mapping also supports the homogeneous elemental distribution.
Table 1 shows the atomic compositions of the samples determined by EDX measurements.
The respective composition agrees with the stating composition.

\begin{figure}
\begin{center}
\includegraphics[width=\linewidth]{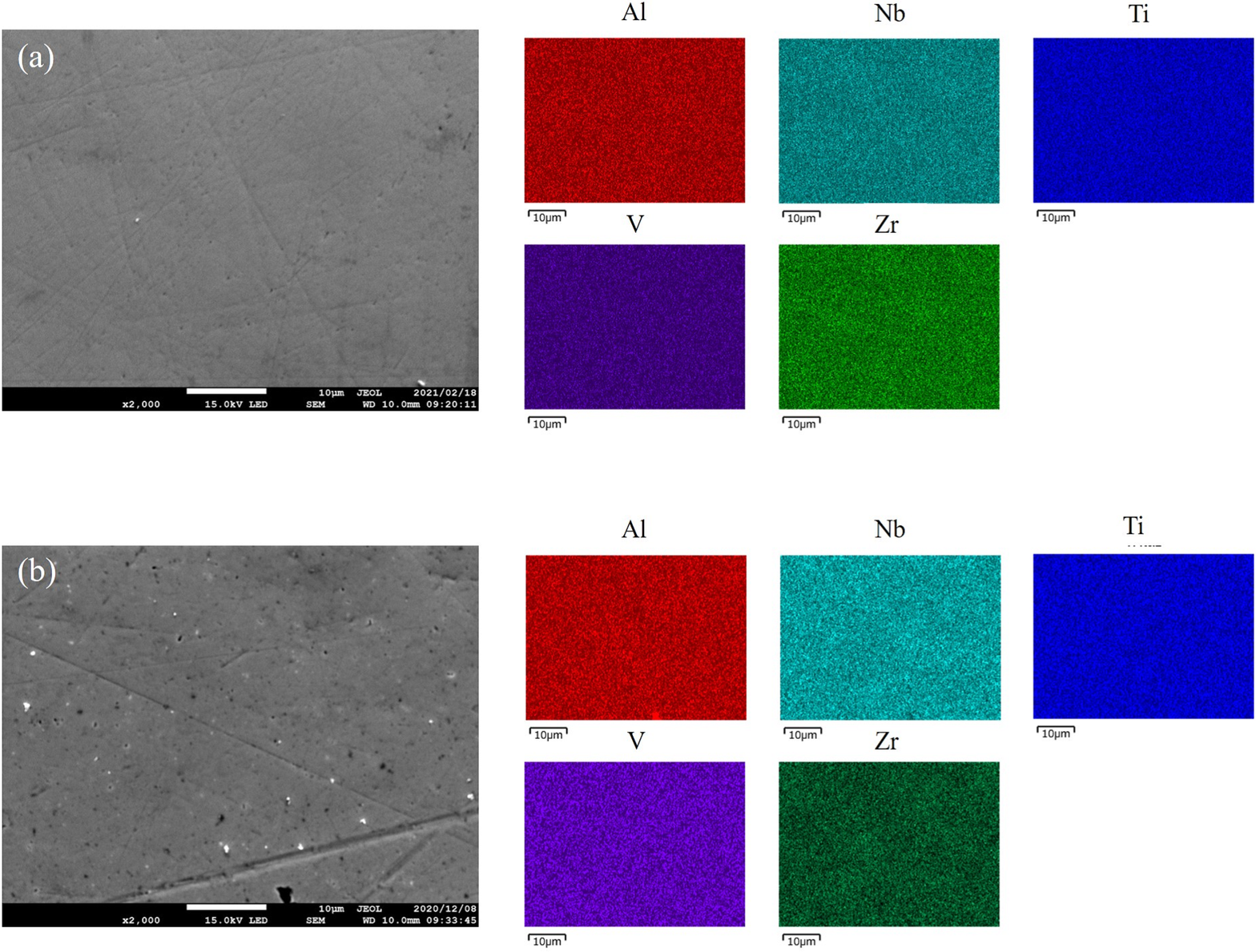}
\end{center}
\caption{SEM images of (a) Al$_{5}$Nb$_{24}$Ti$_{35}$V$_{5}$Zr$_{31}$ and (b) Al$_{5}$Nb$_{34}$Ti$_{25}$V$_{5}$Zr$_{31}$, respectively. The elemental mappings are also shown.}
\label{f3}
\end{figure}

Shown in Figs.\ 4(a) and 4(b) are $\chi_{ac}^{'}$ ($T$) of Al$_{5}$Nb$_{x}$Ti$_{35}$V$_{5}$Zr$_{55-x}$ and Al$_{5}$Nb$_{x}$Ti$_{59-x}$V$_{5}$Zr$_{31}$, respectively.
All samples exhibit diamagnetic signals.
In both alloy systems, $T_\mathrm{c}$ systematically increases with increasing Nb concentration.
Figures 5(a) and 5(b) exhibit $\rho$ ($T$) of Al$_{5}$Nb$_{x}$Ti$_{35}$V$_{5}$Zr$_{55-x}$ and Al$_{5}$Nb$_{x}$Ti$_{59-x}$V$_{5}$Zr$_{31}$, respectively, which are  normalized by the room temperature values listed in Table 2.
The temperature dependence in the normal state tends to be weaker with decreasing Nb content.
In each alloy system, the $x$=14 sample exhibit a negative temperature coefficient of resistivity, suggesting an occurrence of carrier localization.
The insets are $\rho$ ($T$) at low temperatures, all showing drops to zero resistivity.
Table 2 summarizes $T_\mathrm{c}$'s in $\chi_{ac}^{'}$ ($T$) and $\rho$ ($T$) measurements, which are determined by the same methods in Al$_{5}$Nb$_{24}$Ti$_{40}$V$_{5}$Zr$_{26}$.

\begin{figure}
\begin{center}
\includegraphics[width=\linewidth]{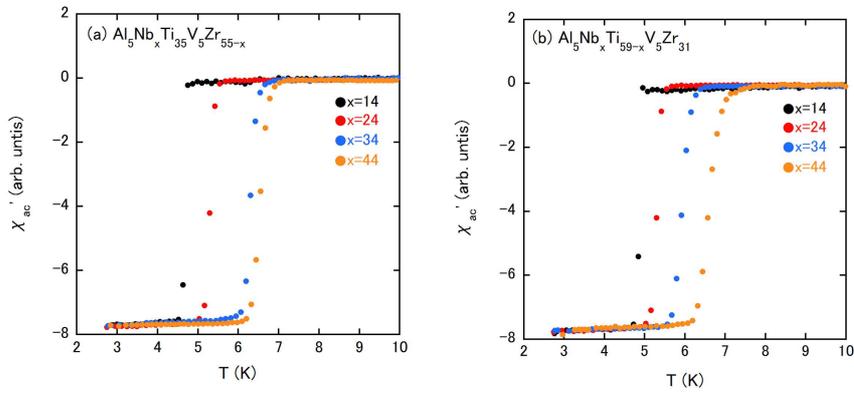}
\end{center}
\caption{Temperature dependences of real part of $\chi_{ac}$ for (a) Al$_{5}$Nb$_{x}$Ti$_{35}$V$_{5}$Zr$_{55-x}$ and (b) Al$_{5}$Nb$_{x}$Ti$_{59-x}$V$_{5}$Zr$_{31}$, respectively.}
\label{f4}
\end{figure}

\begin{figure}
\begin{center}
\includegraphics[width=\linewidth]{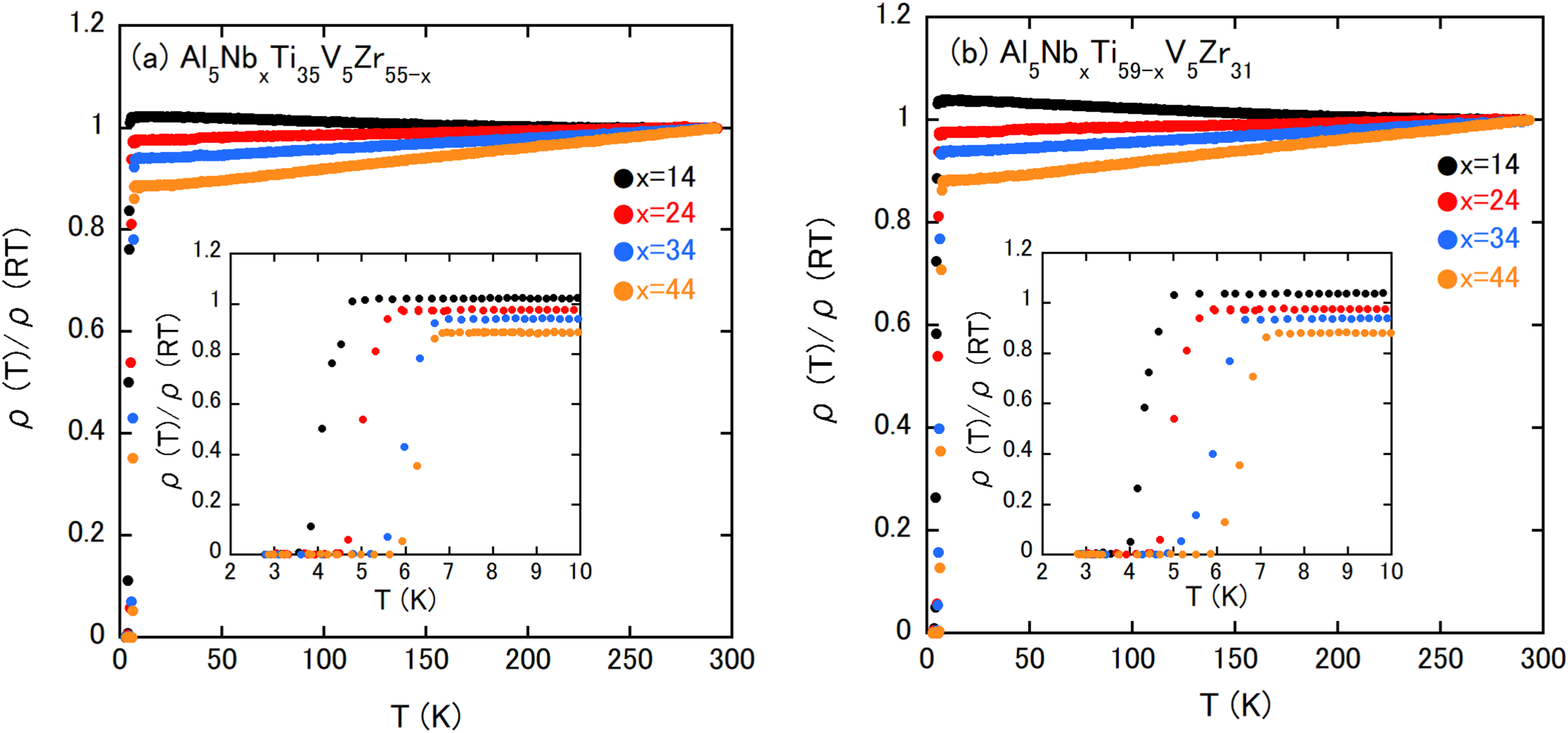}
\end{center}
\caption{Temperature dependences of $\rho$ normalized by room temperature (RT) values for (a) Al$_{5}$Nb$_{x}$Ti$_{35}$V$_{5}$Zr$_{55-x}$ and (b) Al$_{5}$Nb$_{x}$Ti$_{59-x}$V$_{5}$Zr$_{31}$, respectively. The insets are $\rho$ ($T$) at low temperatures.}
\label{f5}
\end{figure}

\begin{table}
\caption{VEC, $T_\mathrm{c}$'s determined by $\chi_{ac}$ and $\rho$ measurements, and $\rho$ at room temperature of prepared samples.}
\label{t2}
\begin{tabular}{ccccc}
\hline
Sample & VEC & $T_\mathrm{c}$ (K) & $T_\mathrm{c}$ (K) & $\rho$ at room   \\
       &     & by $\chi_{ac}$ & by $\rho$ & temperature ($\mu\Omega$cm)   \\
\hline
Al$_{5}$Nb$_{24}$Ti$_{40}$V$_{5}$Zr$_{26}$ & 4.24 & 5.50 & 5.36 & 102.3 \\
Al$_{5}$Nb$_{14}$Ti$_{35}$V$_{5}$Zr$_{41}$ & 4.14 & 4.73 & 4.40 & 88.3 \\
Al$_{5}$Nb$_{24}$Ti$_{35}$V$_{5}$Zr$_{31}$ & 4.24 & 5.45 & 5.43 & 111.5 \\
Al$_{5}$Nb$_{34}$Ti$_{35}$V$_{5}$Zr$_{21}$ & 4.34 & 6.47 & 6.48 & 105.2 \\
Al$_{5}$Nb$_{44}$Ti$_{35}$V$_{5}$Zr$_{11}$ & 4.44 & 6.75 & 6.66 & 69.4 \\
Al$_{5}$Nb$_{14}$Ti$_{45}$V$_{5}$Zr$_{31}$ & 4.14 & 4.96 & 4.65 & 111.6 \\
Al$_{5}$Nb$_{34}$Ti$_{25}$V$_{5}$Zr$_{31}$ & 4.34 & 6.18 & 6.38 & 61.3 \\
Al$_{5}$Nb$_{44}$Ti$_{15}$V$_{5}$Zr$_{31}$ & 4.44 & 6.98 & 7.01 & 76.6 \\
\hline
\end{tabular}
\end{table}

Here we discuss the relationship between $T_\mathrm{c}$ and the parameter $\delta$ or mixing entropy $\Delta S_\mathrm{mix}$.
The parameter $\delta$ is employed in the material design of HEAs and means the degree of the atomic size difference among the constituent elements\cite{Gao:book,Murty:book}.
This was calculated for an $n$-component alloy as follows:
\begin{equation}
\delta=100\times\sqrt{\sum^{n}_{i=1}c_{i}\left(1-\frac{r_{i}}{\bar{r}}\right)^{2}},
\label{equ1}
\end{equation}
where $c_{i}$ and $r_{i}$ are the atomic fraction and the atomic radius of the $i$-th element, respectively, and $\bar{r}$ is the composition-weighted average atomic radius.
The values of $r_{i}$ are taken from ref.\cite{Miracle:ActaMater}.
$\Delta S_\mathrm{mix}$ is given by
\begin{equation}
\Delta S_\mathrm{mix}=-R\sum^{n}_{i=1}c_{i}lnc_{i},
\label{equ2}
\end{equation}
where $R$ is the gas constant.
As mentioned in the introduction, there are two well-accepted definitions\cite{Gao:book,Murty:book} for HEAs.
Although the three alloys Al$_{5}$Nb$_{24}$Ti$_{35}$V$_{5}$Zr$_{31}$, Al$_{5}$Nb$_{34}$Ti$_{35}$V$_{5}$Zr$_{21}$, and Al$_{5}$Nb$_{34}$Ti$_{25}$V$_{5}$Zr$_{31}$ fulfill the definition by the atomic percentage, each having 5 \% $\leq$ $c_{i}$ $\leq$ 35 \%, $\Delta S_\mathrm{mix}$'s of all alloys investigated are rather smaller than 1.60$R$ (see also Table 1).
Figures 6(a) and 6(b) explain the $\delta$ or $\Delta S_\mathrm{mix}/R$ dependences of $T_\mathrm{c}$ for both alloy systems.
For Al$_{5}$Nb$_{x}$Ti$_{35}$V$_{5}$Zr$_{55-x}$, $T_\mathrm{c}$ increases with decreasing $\delta$, which is contrasted with the $\delta$ dependence of $T_\mathrm{c}$ in Al$_{5}$Nb$_{x}$Ti$_{59-x}$V$_{5}$Zr$_{31}$.
This means the robustness of superconductivity against the lattice distortion, which would be one of the characteristics of bcc HEA superconductors.
In both alloy systems, $T_\mathrm{c}$ shows the non-monotonous $\Delta S_\mathrm{mix}$ dependences.
Therefore, $T_\mathrm{c}$ does not severely rely on the high-entropy effect.

\begin{figure}
\begin{center}
\includegraphics[width=\linewidth]{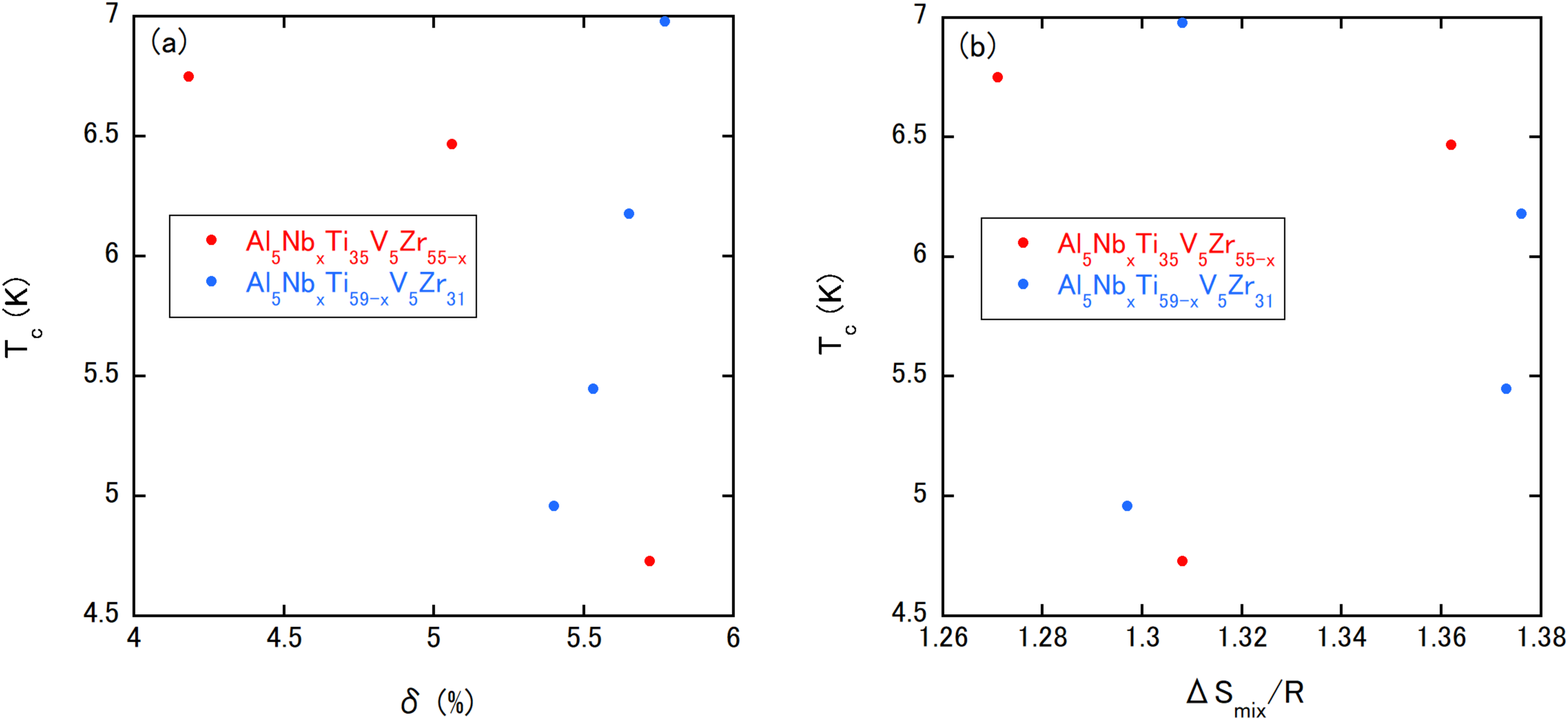}
\end{center}
\caption{(a)$T_\mathrm{c}$ vs $\delta$ plot and (b)$T_\mathrm{c}$ vs $\Delta S_\mathrm{mix}/R$ plot for Al$_{5}$Nb$_{x}$Ti$_{35}$V$_{5}$Zr$_{55-x}$ and Al$_{5}$Nb$_{x}$Ti$_{59-x}$V$_{5}$Zr$_{31}$.}
\label{f6}
\end{figure}

Figure 7 shows the VEC dependence of $T_\mathrm{c}$ for Al-Nb-Ti-V-Zr multicomponent alloys, which is compared to those of crystalline 4$d$ metal solid solutions\cite{Sun:PRM} (the dotted line) and typical superconducting bcc HEAs\cite{Rohr:PNAC,Kozelj:PRL,Marik:JALCOM,Ishizu:RINP} (Ta-Nb-Hf-Zr-Ti, Nb$_{20}$Re$_{20}$Zr$_{20}$Hf$_{20}$Ti$_{20}$, and Hf$_{21}$Nb$_{25}$Ti$_{15}$V$_{15}$Zr$_{24}$).
Although $T_\mathrm{c}$'s of these HEAs are relatively lower than those of 4$d$ metal solid solutions at a fixed VEC, the VEC dependence is similar to each other; $T_\mathrm{c}$ seems to form a broad maximum at approximately VEC=4.7.
The VEC dependence of $T_\mathrm{c}$ of Al-Nb-Ti-V-Zr multicomponent alloy falls on those of bcc HEAs, whereas $\Delta S_\mathrm{mix}$'s of the present alloys are lower.
Ternary alloys Nb-Ti-Zr are also superconductors\cite{Roberts:JPCRD}, and the data is plotted in Figure 7 (Nb$_{50}$Ti$_{10}$Zr$_{40}$: $T_\mathrm{c}=$10.3 K, Nb$_{35}$Ti$_{15}$Zr$_{50}$: $T_\mathrm{c}=$8.95 K, Nb$_{35}$Ti$_{60}$Zr$_{5}$: $T_\mathrm{c}=$8.6 K, and Nb$_{21}$Ti$_{61}$Zr$_{18}$: $T_\mathrm{c}=$6.8 K).
The plot well agrees with the Matthias rule of 4$d$ metal solid solutions.
Therefore, a multicomponent alloy composed of five elements might generally show the lowered $T_\mathrm{c}$ compared to simple 4$d$ metal solid solutions.

\begin{figure}
\begin{center}
\includegraphics[width=0.8\linewidth]{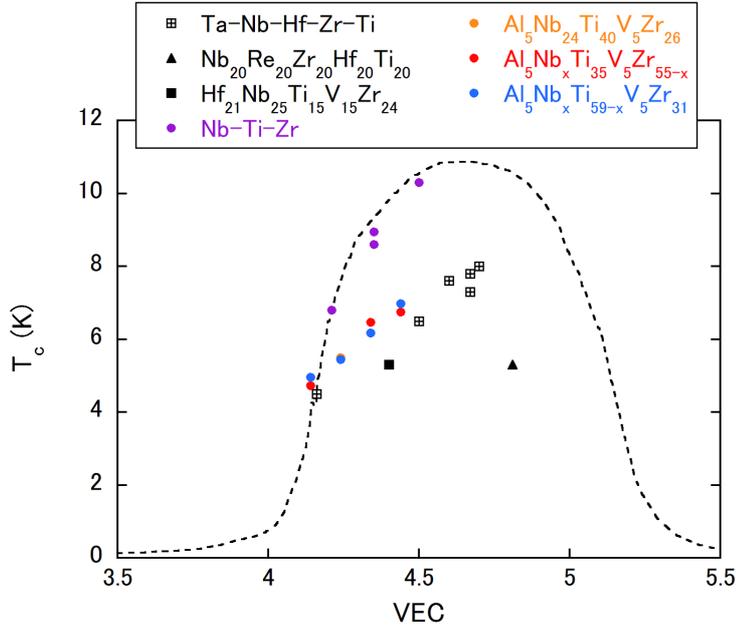}
\end{center}
\caption{VEC dependences of $T_\mathrm{c}$ for alloys denoted in figure legend. The same dependence is also shown for crystalline 4$d$ metal solid solutions by the dotted line. Data for the dotted line is from ref.\cite{Sun:PRM}.}
\label{f8}
\end{figure}

Several superconductors maintain the superconducting state after the cold rolling process\cite{Raabe:ActaMater,Guryev:JPCS,Idczak:PLA}.
Thus we can expect the same result for the present alloys.
It would be important to elucidate the relationship between the mechanical properties of Al-Nb-Ti-V-Zr alloys, including gum-metal composition, and the superconducting properties in order to know whether the gum-metal HEA superconductors can be candidate for the next-generation superconducting wire or not.

\section{Summary}
We have characterized new multicomponent alloy superconductors Al-Nb-Ti-V-Zr including Al$_{5}$Nb$_{24}$Ti$_{40}$V$_{5}$Zr$_{26}$ which exhibits a gum-metal-like behavior after the cold rolling.
All as-cast samples crystallize into the bcc structure with no obvious secondary phase.
The lattice parameters of Al$_{5}$Nb$_{x}$Ti$_{35}$V$_{5}$Zr$_{55-x}$ and Al$_{5}$Nb$_{x}$Ti$_{59-x}$V$_{5}$Zr$_{31}$ exhibit the different $x$ dependences.
Al$_{5}$Nb$_{x}$Ti$_{35}$V$_{5}$Zr$_{55-x}$ shows the shrink of lattice parameter with increasing $x$, whereas the almost $x$-independent lattice parameters are observed in Al$_{5}$Nb$_{x}$Ti$_{59-x}$V$_{5}$Zr$_{31}$.
Nonetheless, the similar VEC dependence of $T_\mathrm{c}$ is confirmed for both alloy systems.
Although the ternary superconducting Nb-Ti-Zr alloys well obey the Matthias rule of crystalline 4$d$ metal solid solutions, the present alloys show lower $T_\mathrm{c}$'s compared to ternary Nb-Ti-Zr alloys.
However, the VEC dependence of $T_\mathrm{c}$ of Al-Nb-Ti-V-Zr alloys falls on those of the typical bcc HEA superconductors.
Therefore, a multicomponent alloy composed of five elements might generally behave as an HEA superconductor, even if the mixing entropy is rather smaller than 1.60$R$.
This study will be useful for the fundamental understanding of superconductivity in gum metals.

\begin{acknowledgements}
J.K. is grateful for the support provided by Comprehensive Research Organization of Fukuoka Institute of Technology.
\end{acknowledgements}

\end{document}